\documentclass[useAMS,usenatbib]{mn2e}

\usepackage{graphicx}
\usepackage{wasysym}
\usepackage{amssymb}
\usepackage{stmaryrd}

\title[Photometric Study and Detection of Variable Stars in the Open Clusters - I : NGC~6866]
      {Photometric Study and Detection of Variable Stars in the Open Clusters - I : NGC~6866}

\author[Y.~C.~Joshi et al.]{Y.~C.~Joshi$^{1}$\thanks{E-mail: yogesh@aries.res.in},
S.~Joshi$^{1}$,
Brijesh~Kumar$^{1}$, 
Soumen~Mondal$^{1,2}$, 
L.~A.~Balona$^{3}$\\
$^{1}$Aryabhatta Research Institute of Observational Sciences (ARIES), Manora peak, Nainital, India	\\
$^{2}$Department of Astrophysics and Cosmology, S.N. Bose National Centre for Basic Sciences, Salt Lake, Kolkata-700 098, India\\
$^{3}$South African Astronomical Observatory, PO Box 9, Observatory 7935, Cape Town, South Africa	\\
}
\begin{document}

\date{Accepted 2011 September 24;  Received 2011 September 21; in original form 2011 July 8}


\maketitle

\label{firstpage}

\begin{abstract}
We present results of a variability search in the 
intermediate-age open cluster NGC~6866 from 29 nights over two observing
seasons. We find 28 periodic variables, of which 19 are newly identified. The 
periods of these variables, which have $V$ magnitudes from 11.5  to 19.3~mag, 
range from $\sim$ 48~min to 37~d. We detected several $\delta$-Scuti stars,
some of which are of high amplitude, as well as $\gamma$-Doradus, rotational 
variables and eclipsing binaries. In order to study the physical properties 
of the cluster, we obtained $UBVRI$ photometry of all the stars on a good 
photometric night. The radial distribution of the stellar surface density 
shows that the cluster has a radial extent of about 7 arcmin ($\sim$ 3 pc)
with a peak density of $5.7\pm0.7$ stars/arcmin$^2$ at the cluster center. 
The colour-colour diagram indicates a reddening of $E(B-V) = 0.10$ mag
towards NGC~6866. A distance of $\sim$1.47 kpc and an age of $\sim$
630 Myr is estimated from the colour-magnitude diagram using the theoretical
isochrones of solar metallicity. 
\end{abstract}


\section{INTRODUCTION} \label{sec:intro}
The study of Galactic open clusters is important for understanding 
the history of star formation and the nature of the parent star clusters. 
These systems are used to test stellar models and are vital for our
understanding of stellar evolution. By comparing the colour-magnitude 
diagram of a star cluster with theoretical evolutionary models, its age may
be estimated and information on the evolution of stars of nearly the same 
age and chemical composition could be obtained. However, uncertainties in various 
parameters such as reddening, distance and chemical composition compromise
our efforts to measure the age and other parameters. Some cluster 
members also show various kinds of variability at different stages of 
their evolution. For example, a wide variety of pulsating variable 
stars such as $\delta$~Scuti and $\beta$~Cephei stars are found in open 
star clusters. Observations of these variables can be used to infer 
many important stellar parameters such as their masses, radii and 
luminosities. These parameters are required for an investigation of 
the relationships between rotation, stellar activity, age and masses
(Messina et al. 2008, 2010) and impose constraints on stellar pulsation 
models. Recent work on cluster variable stars can be found in Hargis \& 
Sandquist~(2005), Parihar et al.~(2009), Marchi et al.~(2010), Saesen et 
al.~(2010), among others.

At ARIES, Nainital, we have started a long-term observational survey
of variable stars in unstudied or poorly studied young and 
intermediate age open clusters. The purpose of this survey is to search 
and characterize the variable star content of these clusters and to
determine their fundamental parameters. Young clusters allow the study 
of pre-main sequence (PMS) stars and the effects of variability on the 
spread in cluster ages. Intermediate age  clusters can be used to study 
short-period variables such as $\delta$~Scuti and $\gamma$~Dor stars.
Intensive un-interrupted observations over a few hours, 
as well as extensive observations spanning several months and years, are
planned for many such clusters.

In this paper, we present a study of the intermediate age open cluster 
NGC~6866 (RA = 20:03:55, DEC = +44:09:30; $l = 79^\circ.560, b = +6^\circ.839$). 
The cluster is in the field observed by the {\it Kepler} satellite, whose
aim is to search for earth-like planets by monitoring the light variations
of more than 150\,000 stars in a 105 square degree field situated in the 
Cygnus-Lyra  region. The data obtained in this mission are also used for the 
asteroseismic study of pulsating variables and to investigate stellar 
characteristics with micro-magnitude precision and uninterrupted time coverage 
(see, Basri et al. 2010). The cluster NGC~6866 has not been extensively 
surveyed for variability. The Hipparcos and Tycho catalogues contain a 
few suspected variables in the cluster. Molenda-Zakowicz et al. (2009) 
obtained time-series CCD photometric observations of this cluster for 14 
nights over a period of about 3 months. They found 19 variables in the field 
of the cluster, of which 12 seem to be  periodic variables. However, due to 
their limited time coverage, the periods and characteristics of these
variables are somewhat uncertain.

Details of the observations and data reduction are presented in \textsection 2. 
The photometric calibration and cluster parameters are studied in \textsection 3.
In \textsection 4 we discuss identification of variable stars and their 
characteristics. This is followed by a discussion and the conclusions.

\section{Observations and data reduction} \label{sec:obs}
Time-series observations of stars in NGC~6866 were obtained in
the Johnson $V$ and Cousin $I$ bands using the 1.04-m Sampurnanand telescope 
at Manora Peak, Nainital. We used a 2k $\times$ 2k CCD camera covering a 
field of view of about 13 $\times$ 13 arcmin. The readout noise and gain of 
the CCD are 5.3~e- and 10~e-/ADU respectively. The cluster was observed for 
29 nights between 2008 September 26 and 2011 January 10 over two observing 
seasons. Priority was given to $V$-band observations in order to increase 
the time sampling in this waveband. A total of 768 frames in 
the $V$ band and 50 frames in the $I$ band were accumulated. Bias 
and twilight flat frames were taken on a regular basis. To improve the 
signal-to-noise ratio, all observations were taken in $2 \times 2$ binning 
mode ($\approx 0.74\times0.74$~arcmin). The exposure time for each frame 
ranged from 30~sec to 200~sec depending on sky conditions and time constraints. 
Typical seeing was $\sim$ 2 arcsec. In order to detect short-period variables
we observed NGC~6866 continuously for more than 4 hours on 3 different nights 
with an observing cadence of about 2.2~min. A brief log of the observations is 
given in Table 1.
\begin{table}
  \caption{Log of the CCD observations.}
  \label{tab:obslog}

  \begin{tabular} {cccc}
  \hline
  Date & JDstart & Filters & No. of frames \\ 
       &(+2450000) &        &           \\ \hline
2008 09 26  &   4736.088461 &  V   &   46  \\
2008 09 27  &   4737.071076 &  V,I   &   57,2\\
2008 09 28  &   4738.090602 &  V   &   54  \\
2008 09 30  &   4740.084549 &  V   &   50  \\
2010 09 30  &   5470.132100 &  V,I &    3,3\\
2010 10 05  &   5475.112778 &  V   &   51  \\
2010 10 06  &   5476.106863 &  V   &  100  \\
2010 10 07  &   5477.060637 &  V   &  142  \\
2010 10 08  &   5478.046644 &  V   &    5  \\
2010 10 10  &   5480.104676 &  V   &  108  \\
2010 10 12  &   5482.275231 &  V   &    3  \\
2010 10 20  &   5490.058611 &  V   &    3  \\
2010 10 24  &   5494.084444 &  V   &    3  \\
2010 10 25  &   5495.117824 &  V   &    3  \\
2010 10 26  &   5496.076435 &  V,I &    3,3\\
2010 10 27  &   5497.068206 &    I &       3\\
2010 10 28  &   5498.065243 &  V,I &    3,3\\
2010 10 31  &   5501.105200 &  V,I &    3,3\\
2010 11 01  &   5502.088067 &  V,I &    3,3\\
2010 11 02  &   5503.081900 &  V,I &    7,3\\
2010 11 03  &   5504.048300 &  V,I &   61,3\\
2010 11 04  &   5505.111852 &  V   &   51  \\
2010 11 30  &   5531.032500 &  V,I &    2,2\\
2010 12 02  &   5533.042700 &  V,I &    2,2\\
2010 12 24  &   5555.034086 &  V,I &    2,2\\
2011 01 04  &   5566.040800 &  I   &    3  \\
2011 01 05  &   5567.041516 &  V,I &    3,6\\
2011 01 06  &   5568.051100 &  I   &    3  \\
2011 01 10  &   5572.043906 &  I   &    6  \\
\hline
  \end{tabular}                                                                     

\end{table} 

The basic steps of image processing, which include bias subtraction, flat 
fielding, and cosmic ray removal, were performed using IRAF. During the 
observations, it was not possible to keep all the stars at the same pixel 
position on every night. The IRAF tasks GEOMAP and GEOTRAN were used to 
align all the images with respect to a reference frame which was chosen so
that the cluster center falls approximately at the center of the observed 
field. In most cases, we achieved a pixel accuracy of $\approx$~0.1 arcsec.
\section{Photometric study of the cluster NGC~6866} \label{sec:phot} 
\subsection{Photometric Calibration} \label{sec:photcal}
Photometry was performed using DAOPHOT II profile fitting software (Stetson 1987).
In order to perform consistent photometry from night to night on the aligned images,
we made a master list of 2809 stars from 13 frames which have the best seeing and
coverage of the target field. It should be noted that not all stars could be measured
on every frame owing to different exposure times and sky conditions.

To determine the extinction coefficients, colour terms and zero points, we
used Landolt's (1992) standard fields SA95 and  PG 0231$+$051. These were
observed on the night of 2010 November 30 in photometric sky conditions.
Exposure times of 300, 300, 200, 100, and 60-sec were used for $U$, $B$, $V$, $R$
and $I$ respectively. We acquired two frames each of NGC~6866 and PG 0231+051 and
four frames of SA95. A total of fifteen standard stars with $12.68 < V < 16.28$ and 
$-0.53 < (V-I) < 1.95$ were observed during this night with airmasses ranging from 
1.11 to 2.04. Nine of these stars are in SA95 and six in PG 0231$+$051.
We used profile-fitting photometry to extract the magnitudes using the program 
{\it DAOGROW} for obtaining the curve-of-growth corrections. A  linear
least square regression was used to the standard system to derive the following
transformation equations:\\
\\
   $ u = U + 8.13\pm0.01 - (0.06\pm0.01)(U-B) + (0.59\pm0.04)X $ \\
   $ b = B + 5.84\pm0.02 - (0.02\pm0.02)(B-V) + (0.27\pm0.03)X $\\
   $ v = V + 5.40\pm0.01 - (0.02\pm0.01)(B-V) + (0.17\pm0.02)X $\\
   $ r = R + 5.22\pm0.01 - (0.04\pm0.02)(V-R) + (0.09\pm0.02)X $\\
   $ i = I + 5.63\pm0.02 + (0.01\pm0.01)(R-I) + (0.05\pm0.02)X $\\
   
\noindent
where $u, b, v, r$ and $i$ are the aperture instrumental magnitudes
and $U, B, V, R$ and $I$ are the standard magnitudes and $X$ is
the airmass. Zero-point and colour-coefficient errors are $\sim$ 0.01 mag.
These equations were used to generate secondary standard stars in the target 
field observed on the same night. To standardize the data on remaining nights, 
differential photometry was performed using these secondary stars. We used
a linear fit between the standard and instrumental magnitudes on each night,
assuming that most of the stars are non-variables. We rejected stars which
deviated by more than $3-\sigma$ deviations. 
In Fig.~1, we show the standard deviations as a function of magnitudes
in $UBVRI$ bands. The photometric errors were computed on the night of 
standardization. The typical standard deviation is less than 0.05 mag for stars
brighter than $\sim$ 19 mag in $V$ band. 

\begin{figure} 
\includegraphics[height=12cm,width=8cm]{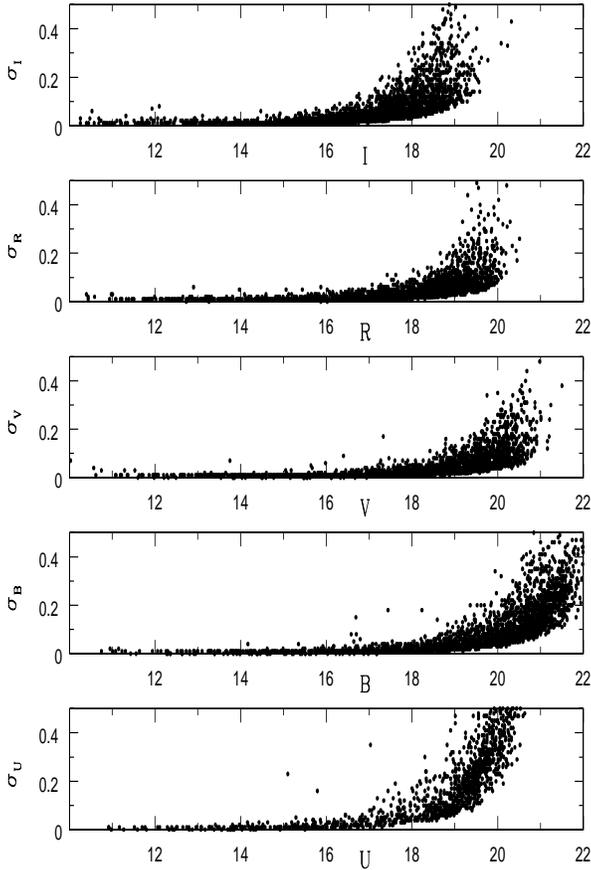} 
\caption{The photometric error of stars as a function of brightness in $UBVRI$ bands.}
\label{fig:comp} 
\end{figure}
%
%
\subsection{Comparison with previous photometry} \label{sec:photcomp} 
NGC~6866 has not been intensively observed in the past. $UBV$ photoelectric
and photographic photometry were obtained by Hoag et al. (1961), Johnson et al. 
(1961), and Barkhatova and Zakharova (1970), mostly for bright stars. Recently, 
Molenda-Zakowicz et al. (2009; hereafter MOL09) obtained $BVI_C$ CCD
photometry in a $12.8\times11.7$ arcmin field centred on NGC~6866 {\bf using the 60-cm 
telescope at the Bial\'{k}ow Observatory of the Wroclaw University, Poland.}
They observed 552 stars and derived calibrated $B$ and $V$ photometry.
We did not compare our photometry with the older photoelectric and photographic 
work as these contain relatively large errors and are mostly confined to 
the brighter stars. We do, however, compare our results with MOL09 which,
in turn, have been compared with older photoelectric and photographic
measurements. We have cross-identified our stars with those of MOL09 and
find 511 stars in common. Fig.~2 shows the difference, $\Delta$, in
$V$ and $(B-V)$ between our photometry and MOL09.
A linear fit between our $V$ magnitudes and those of MOL09 shows a magnitude 
dependence which can be given by the following equation
$$ V_{present} - V_{Mol} = (0.008\pm0.001) \times V_{present} - (0.13\pm0.02) $$
\begin{figure} 
\includegraphics[width=8cm]{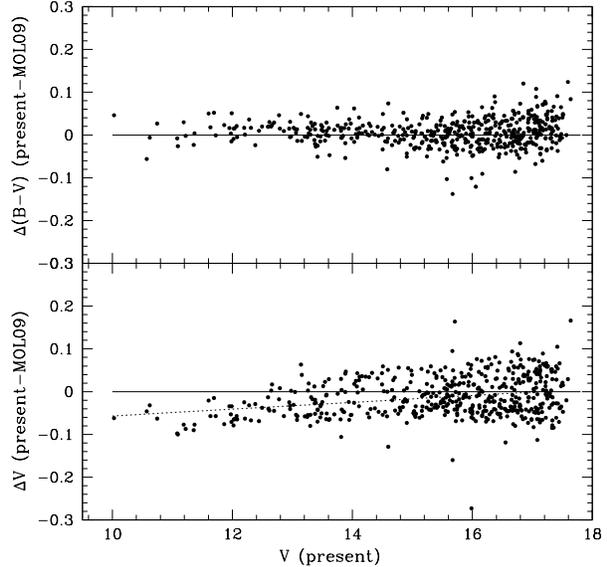} 
\caption{A comparison of our $V$ and $(B-V)$ observations with those of
MOL09. The solid line represents a zero magnitude difference while the dashed line 
in the lower panel is a least-squares linear fit.}
\label{fig:comp} 
\end{figure}
%
{\bf Our $V$ magnitudes are noticeably brighter than MOL09 for the brightest stars,
particularly towards $V < 13$ mag.} For faint stars, the $V$ magnitudes and $(B-V)$
colours between two photometry seem to be in agreement within their internal
photometric errors.
\subsection{Astrometry} \label{sec:photast}
To convert the pixel coordinates (X,Y) into
celestial coordinates ($\alpha_{2000}$, $\delta_{2000}$), a linear
astrometric solution was derived by matching about 134 bright stars in common 
between our catalogue and the 2MASS catalogue. We used the {\it CCMAP} and 
{\it CCTRAN} tasks inside IRAF for this purpose. Using these equations, we converted 
the (X,Y) position into J2000 celestial coordinates for all the stars in our
catalogue. The typical rms scatter, as derived from the difference between 
calibrated and catalogued coordinates, is about 0.1~arcsec. 


\begin{table*}
\caption{UBVRI photometric catalogue of 2473 stars detected in the field of cluster NGC~6866.
The error in magnitudes indicates the internal photometric error in the measurement. Table
is sorted in the order of increasing $V$ magnitude. Last 3 columns give the membership
probabilities based on the spatial distribution, their position in the CMD and proper motion.}
\label{tab:photclus}
\centering
\begin{tabular}{ccccccccccc}
\hline
ID &    RA      &     DEC    &         U        &        B         &         V       &        R          &         I   & $P_{sp}$& $P_{ph}$& $P_{pm}$      \\
\hline
   1 & 20:04:06.21& 44:04:55.0& 11.70$\pm$0.01 & 11.01$\pm$0.01 & 10.03$\pm$0.07 &  9.54$\pm$0.08 &  9.13$\pm$0.03 & 0.26 & 1.00 & 0.00\\
   2 & 20:03:52.44& 44:11:31.0& 10.92$\pm$0.01 & 10.75$\pm$0.01 & 10.57$\pm$0.04 & 10.40$\pm$0.03 & 10.26$\pm$0.03 & 0.72 & 1.00 & 0.76\\
   3 & 20:04:09.32& 44:04:16.2& 10.98$\pm$0.01 & 10.95$\pm$0.02 & 10.62$\pm$0.01 & 10.44$\pm$0.01 & 10.26$\pm$0.01 & 0.14 & 1.00 & 0.00\\
   4 & 20:03:57.62& 44:08:37.5& 11.27$\pm$0.01 & 11.08$\pm$0.01 & 10.75$\pm$0.03 & 10.59$\pm$0.02 & 10.40$\pm$0.01 & 0.83 & 1.00 & 0.91\\
   5 & 20:04:02.85& 44:11:55.4& 11.52$\pm$0.01 & 11.31$\pm$0.01 & 11.08$\pm$0.01 & 10.94$\pm$0.01 & 10.80$\pm$0.01 & 0.60 & 1.00 & 0.82\\
   6 & 20:03:52.57& 44:11:33.1& 11.17$\pm$0.01 & 11.15$\pm$0.02 & 11.09$\pm$0.01 & 11.01$\pm$0.03 & 11.01$\pm$0.01 & 0.72 & 1.00 & 0.76\\
   7 & 20:03:57.35& 44:09:33.5& 11.59$\pm$0.01 & 11.43$\pm$0.01 & 11.19$\pm$0.01 & 11.08$\pm$0.01 & 10.97$\pm$0.01 & 0.91 & 1.00 & 0.89\\
   8 & 20:03:40.66& 44:11:43.5& 10.92$\pm$0.01 & 11.23$\pm$0.01 & 11.22$\pm$0.01 & 11.20$\pm$0.01 & 11.22$\pm$0.01 & 0.54 & 1.00 & 0.94\\
   9 & 20:03:52.57& 44:15:46.9& 14.90$\pm$0.01 & 12.90$\pm$0.01 & 11.29$\pm$0.03 & 10.44$\pm$0.02 &  9.55$\pm$0.03 & 0.12 & 0.00 & 0.65\\
  10 & 20:04:22.90& 44:10:31.2& 11.77$\pm$0.01 & 11.56$\pm$0.01 & 11.35$\pm$0.01 & 11.23$\pm$0.01 & 11.11$\pm$0.02 & 0.25 & 1.00 & 0.64\\
  11 & 20:03:54.70& 44:10:52.9& 11.80$\pm$0.01 & 11.64$\pm$0.01 & 11.37$\pm$0.01 & 11.21$\pm$0.01 & 11.06$\pm$0.01 & 0.82 & 1.00 & 0.54\\
  12 & 20:03:22.82& 44:15:50.3& 13.29$\pm$0.01 & 12.52$\pm$0.01 & 11.53$\pm$0.03 & 10.98$\pm$0.03 & 10.53$\pm$0.06 & 0.00 & 1.00 & 0.88\\
  13 & 20:03:55.17& 44:08:24.2& 13.26$\pm$0.01 & 12.64$\pm$0.01 & 11.60$\pm$0.01 & 11.09$\pm$0.01 & 10.58$\pm$0.01 & 0.83 & 1.00 & 0.73\\
  14 & 20:03:51.69& 44:10:19.1& 12.06$\pm$0.01 & 11.94$\pm$0.01 & 11.63$\pm$0.01 & 11.48$\pm$0.01 & 11.31$\pm$0.01 & 0.88 & 1.00 & 0.68\\
  15 & 20:03:57.70& 44:06:27.6& 13.38$\pm$0.01 & 12.73$\pm$0.01 & 11.69$\pm$0.01 & 11.19$\pm$0.01 & 10.69$\pm$0.01 & 0.54 & 1.00 & 0.26\\
  16 & 20:03:26.12& 44:10:05.3& 12.11$\pm$0.01 & 12.06$\pm$0.01 & 11.72$\pm$0.01 & 11.51$\pm$0.01 & 11.32$\pm$0.04 & 0.28 & 1.00 & 0.66\\
  17 & 20:03:55.18& 44:10:44.0& 12.29$\pm$0.01 & 12.14$\pm$0.01 & 11.86$\pm$0.01 & 11.74$\pm$0.01 & 11.61$\pm$0.01 & 0.84 & 1.00 & 0.53\\
  18 & 20:03:53.15& 44:09:20.4& 12.33$\pm$0.01 & 12.36$\pm$0.01 & 11.95$\pm$0.01 & 11.74$\pm$0.01 & 11.51$\pm$0.02 & 0.96 & 1.00 & 0.13\\
  19 & 20:04:27.27& 44:10:48.8& 14.03$\pm$0.01 & 13.10$\pm$0.01 & 11.97$\pm$0.01 & 11.39$\pm$0.01 & 10.81$\pm$0.01 & 0.13 & 0.00 & 0.65\\
  20 & 20:04:25.52& 44:10:16.2& 12.45$\pm$0.01 & 12.30$\pm$0.01 & 11.98$\pm$0.01 & 11.81$\pm$0.01 & 11.60$\pm$0.01 & 0.19 & 1.00 & 0.16\\
  21 & 20:03:23.96& 44:10:46.8& 14.49$\pm$0.01 & 13.27$\pm$0.01 & 11.99$\pm$0.01 & 11.33$\pm$0.01 & 10.69$\pm$0.03 & 0.21 & 0.00 & 0.72\\
  22 & 20:03:56.83& 44:10:32.0& 12.41$\pm$0.01 & 12.27$\pm$0.01 & 12.02$\pm$0.01 & 11.91$\pm$0.01 & 11.78$\pm$0.01 & 0.85 & 1.00 & 0.59\\
  23 & 20:03:39.17& 44:14:19.1& 12.50$\pm$0.01 & 12.30$\pm$0.01 & 12.02$\pm$0.01 & 11.85$\pm$0.01 & 11.70$\pm$0.02 & 0.23 & 1.00 & 0.93\\
  24 & 20:03:49.52& 44:10:50.7& 12.39$\pm$0.01 & 12.26$\pm$0.01 & 12.04$\pm$0.01 & 11.94$\pm$0.01 & 11.82$\pm$0.02 & 0.79 & 1.00 & 0.74\\
  25 & 20:04:06.14& 44:12:54.5& 14.22$\pm$0.01 & 13.24$\pm$0.01 & 12.07$\pm$0.01 & 11.46$\pm$0.01 & 10.88$\pm$0.01 & 0.43 & 0.00 & 0.02\\
  26 & 20:04:18.38& 44:09:52.2& 12.52$\pm$0.01 & 12.36$\pm$0.01 & 12.08$\pm$0.01 & 11.91$\pm$0.01 & 11.73$\pm$0.01 & 0.37 & 1.00 & 0.40\\
  27 & 20:03:47.13& 44:09:25.7& 12.65$\pm$0.01 & 12.52$\pm$0.01 & 12.20$\pm$0.01 & 12.02$\pm$0.01 & 11.82$\pm$0.01 & 0.82 & 1.00 & 0.80\\
  28 & 20:04:14.16& 44:12:32.9& 12.87$\pm$0.01 & 12.81$\pm$0.01 & 12.21$\pm$0.01 & 11.87$\pm$0.01 & 11.54$\pm$0.01 & 0.33 & 1.00 & 0.00\\
  29 & 20:03:28.54& 44:09:42.1& 14.36$\pm$0.01 & 13.39$\pm$0.01 & 12.28$\pm$0.01 & 11.72$\pm$0.01 & 11.19$\pm$0.02 & 0.35 & 0.00 & 0.02\\
  30 & 20:04:03.97& 44:10:20.5& 12.74$\pm$0.01 & 12.63$\pm$0.01 & 12.31$\pm$0.01 & 12.14$\pm$0.01 & 11.95$\pm$0.01 & 0.72 & 1.00 & 0.83\\
     &            &           &                &               &               &               &               &        &      &      \\
2472 & 20 03 33.64& 44 13 43.2& - & 21.85$\pm$0.31 & 21.24$\pm$0.30 & 20.19$\pm$0.20 & 18.99$\pm$0.32 & 0.21 & 0.08 & - \\
2473 & 20:03:51.71& 44:10:37.0& - & 22.02$\pm$0.49 & 21.50$\pm$0.38 & 20.51$\pm$0.26 & 19.53$\pm$0.19 & 0.84 & 0.22 & - \\
\hline
\end{tabular}
\end{table*}

In Table~2, we present $UBVRI$ photometric measurements for 2473 stars
identified in the field of the cluster NGC~6866. This is only a sample
listing; the full table is available in electronic form. The table contains 
the star ID, its celestial coordinates for J2000 and $UBVRI$ magnitudes and
their corresponding errors. It should be noted that we were able to
determine the $U$ magnitudes for only about half of these stars owing to the
low quantum efficiency of the CCD in the $U$ band. 
\subsection{Cluster parameters} \label{photpar}
\subsubsection{Radius}
The spatial structure of an open cluster, as derived from the stellar density 
distribution obtained by star counts, is difficult to estimate because of
the irregular shape of the cluster. However, we can obtain some information
about the core and the approximate cluster radius. To  determine the radial 
density distribution (RDP), we divided the cluster into a number of 
concentric annular regions centred on the cluster. The center of the cluster 
was defined by the pixel coordinate (483, 519) in our reference CCD frame. The 
cluster radius is defined as the radius where the surface density becomes
approximately equal to the average density of the surrounding field. To reduce
field star contamination, we only used stars brighter than 19 mag in $V$ in
determining the stellar surface density. The spacing of the annular regions 
was set to 40 pixels ($\sim$ 30 arcsec). This was chosen in such a way that 
each zone contains a statistically significant number of stars.

We counted the number of stars in each annular region. The number density 
in $i^{th}$ zone is then $\rho_i = \frac{N_i}{A_i}$, where $N_i$ and $A_i$ are 
the number of stars and the area of the $i^{th}$ zone respectively. The error
bars are derived assuming that the number of stars in a zone follows Poisson 
statistics. The radial variation of the derived density distribution
is shown in Fig.~3.
\begin{figure} 
\centering 
\includegraphics[width=8cm]{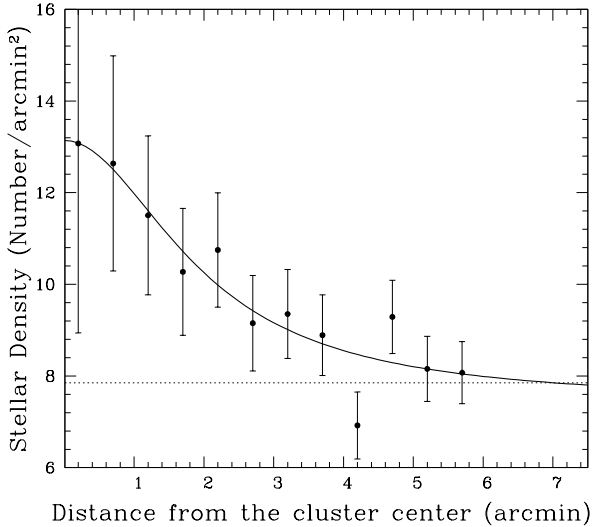} 
\vspace{-1cm}
\caption{The stellar radial density distribution for stars brighter than 
19.0~mag in the field of NGC~6866. The solid line represents the King profile
and the horizontal dashed line shows the average field star density.}
\label{fig:rdp} 
\end{figure} 

Kaluzny (1992) models the projected stellar density distribution as
\begin{center}
$\rho(r) = \rho_f + \frac{\rho_0}{1+ (\frac{r}{r_c})^2} $
\end{center}
\noindent 
where $\rho_0$ is the density at the cluster center, $\rho_f$ is the field
density and $r_c$ is the core radius of cluster. The core radius is defined
as the radius where the density, $\rho(r)$, is half of the central density, 
$\rho_0$. A fit to this function is shown by the solid line in Fig~3.
Using $\chi^2$ minimization, we determined the core radius and central
density to be $2.0\pm0.5$ arcmin and $5.7\pm0.7$ stars/arcmin$^2$ respectively.
Even though our CCD frames did not encompass a large area, we are able to
determine that the surface density is approximately equal to the density
of the surrounding field stars at a radius of about 7 arcmin, which we take as
the cluster radius.
\subsubsection{Reddening}
The reddening, $E(B-V)$, in the cluster region can be determined  using the 
$(U-B)$ vs $(B-V)$ two-colour diagram (TCD). Because only half the stars
have $U$ measurements, and also to avoid contamination by field stars, we
used only stars with  $V<17$ mag and $\sigma_U$, $\sigma_B$, $\sigma_V$ smaller
than 0.05 mag. The resulting TCD is shown in Fig.~4. The observed cluster 
sequence is clearly seen, as is the turn off which occurs at about
spectral type A0. The most probable cluster members with $V < 17.0$~mag and 
$(B-V) < 0.9$~mag are shown by a different symbol.

We adopted the slope of the reddening vector $\frac{E(U-B)}{E(B-V)} = 0.72$. 
We then fitted the intrinsic zero-age main sequence (ZAMS) with solar
metallicity, as given by Schmidt-Kaler (1982), to the observed main-sequence 
(MS) stars. This was done by shifting $E(B-V)$ and $E(U-B)$ along the reddening 
vector (shown in Fig.~4 by the dashed line). The best match results in
a mean reddening of $E(B-V) = 0.10$ mag for NGC~6866. 
%
\begin{figure} 
\includegraphics[width=8cm]{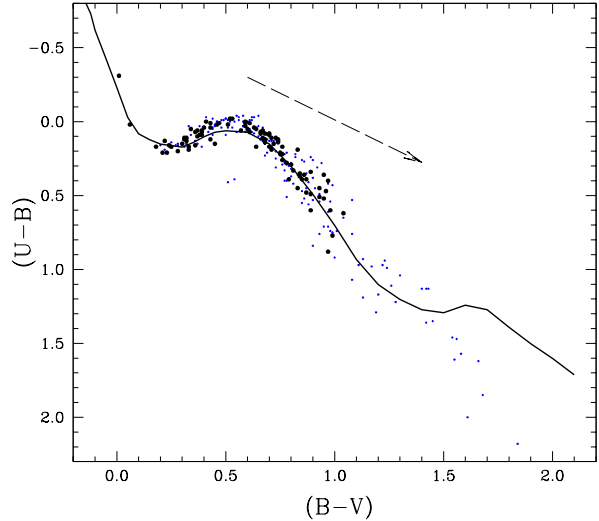} 
\vspace{-1cm}
\caption{The $(U-B)$ versus $(B-V)$ diagram for stars in NGC~6866. The
larger filled circles are stars which lie on or close to the main sequence in 
the colour-magnitude diagram and which have proper motions suggesting more
than a 60 percent probability of being cluster members. The dashed straight 
line represents the slope (0.72) and direction of the reddening vector. The 
solid curve represents the zero-age main sequence from Schmidt-Kaler (1982).
} 
\label{fig:tcd} 
\end{figure}

\subsubsection{Distance and Age}
The identification of the cluster main sequence on the colour-magnitude 
diagrams (CMDs) allows a model-dependent mass, radius, and distance for each star 
to be determined. These are determined by fitting isochrones assuming that
the star is on the  main sequence. To derive these parameters, we
constructed calibrated $(B-V)$, $(V-R)$, and $(V-I)$ vs $V$ diagrams of NGC~6866 
using our data (Fig.~5). A well-defined cluster main sequence is clearly seen in 
all the CMDs. However, they are contaminated by the typical red field star population. 
In order to determine the age and distance of the cluster, theoretical isochrones 
of Girardi et al. (2002) for solar metallicity are over-plotted on the CMDs
(solid lines in Fig.~5). Girardi isochrones are visually fitted by varying the distance
modulus and age simultaneously while
keeping the reddening fixed at $E(B-V) = 0.10$~mag (as estimated earlier using
TCD). We have assumed that the total extinction $A_V = 3.1 \times E(B-V)$. The
best-fit model of the cluster gives a reddening-free distance modulus 
$(V_0-M_V)$ = 10.84~mag and an age of $\log(t) = 8.8$~yrs.

\begin{figure}
\includegraphics[width=8.3cm,height=7.3cm]{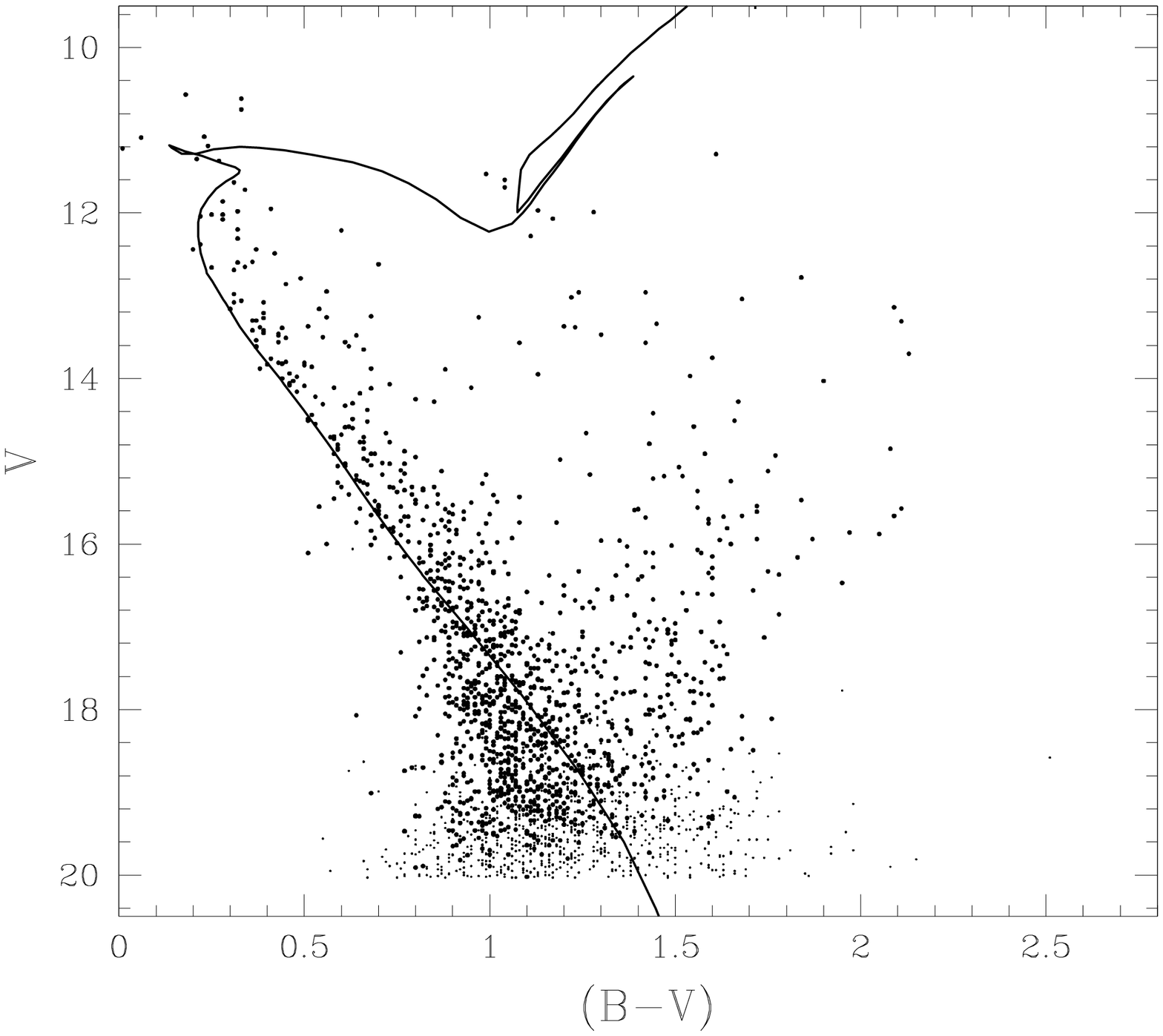}
\includegraphics[width=8.3cm,height=7.3cm]{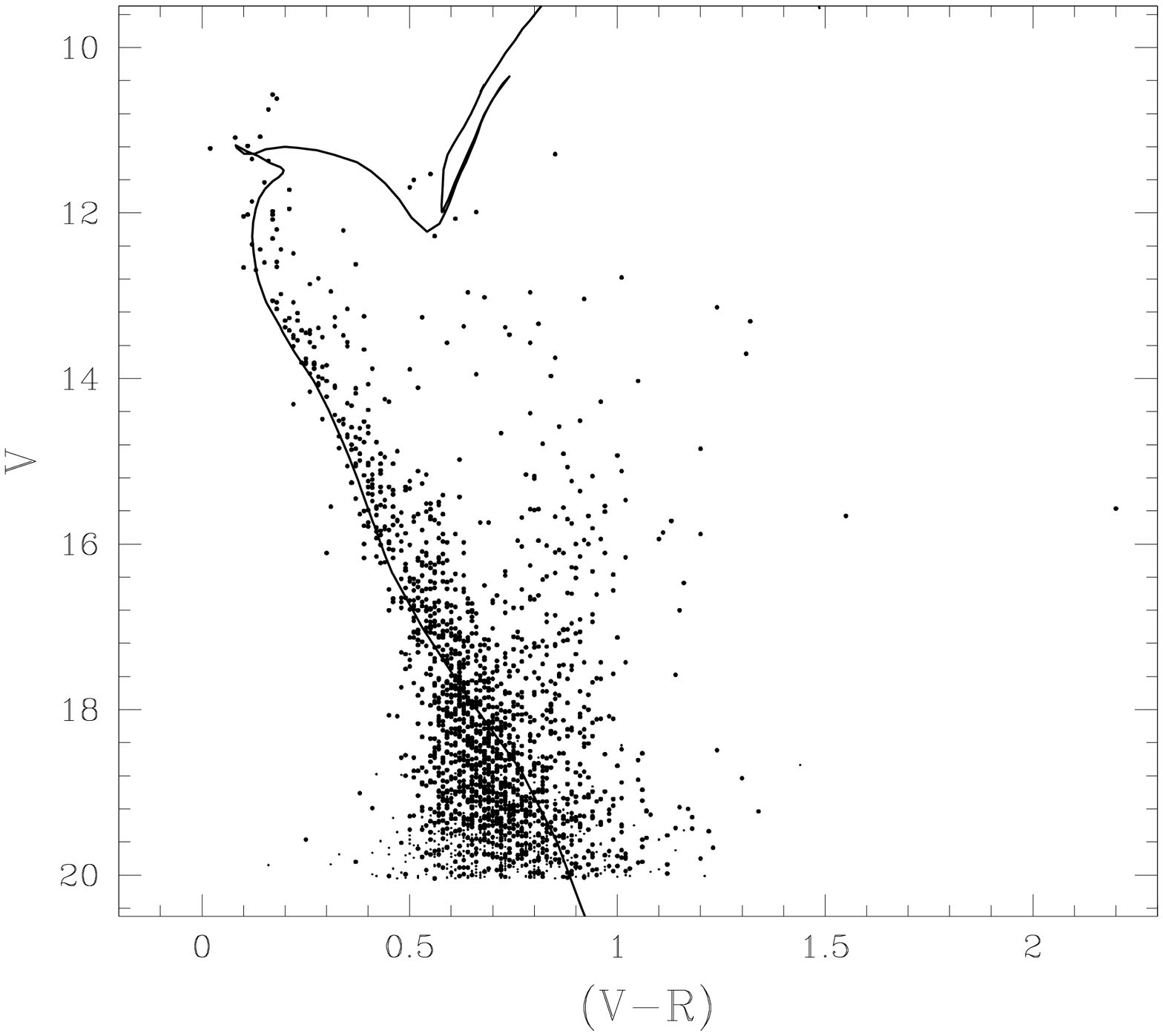}
\includegraphics[width=8.3cm,height=7.3cm]{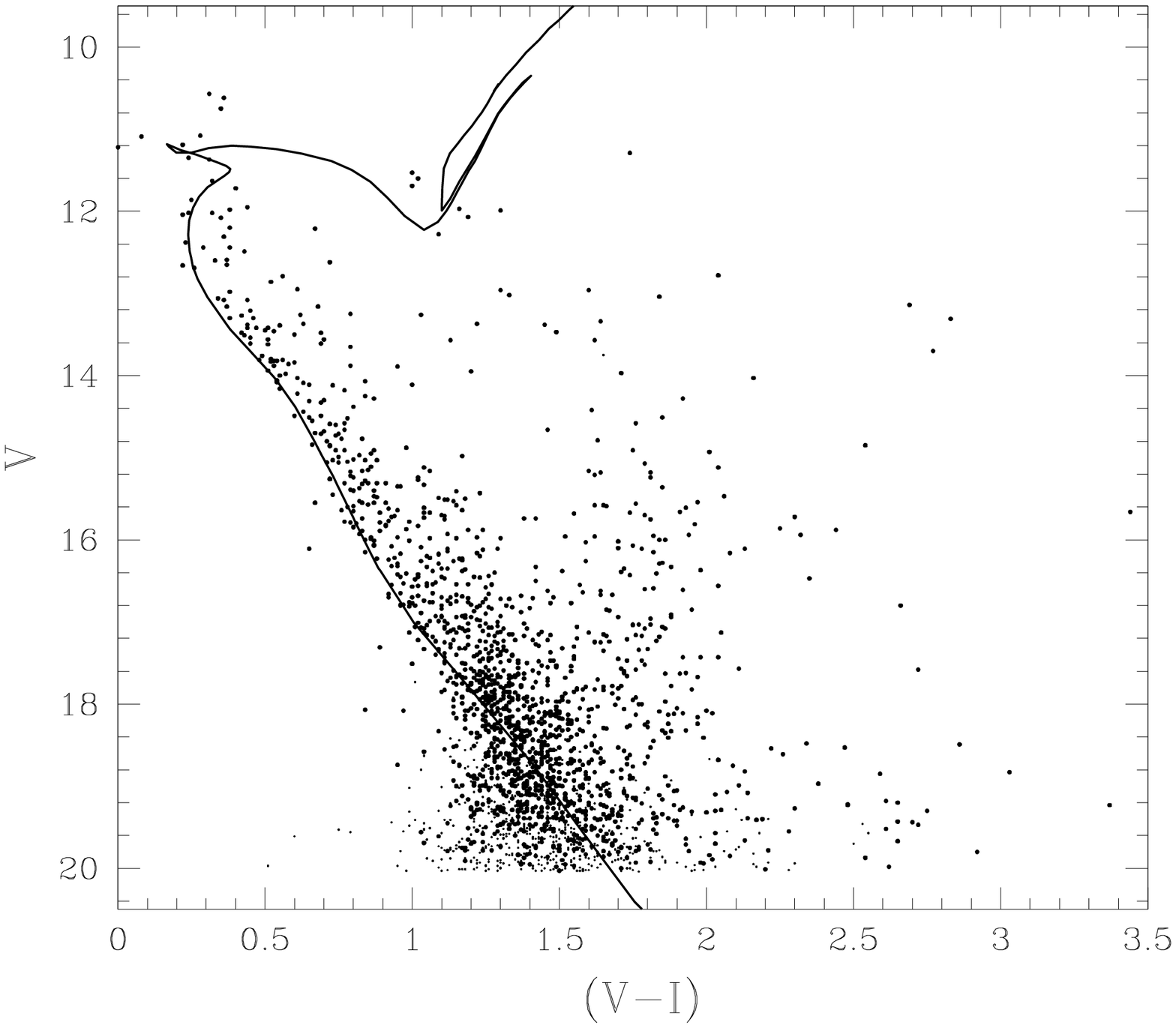} 
\vspace{-0.5cm}
\caption{The colour-magnitude diagrams, $(B-V)$, $(V-R)$ and $(V-I)$ as a function 
of $V$, for stars in the field of NGC~6866. The solid line is the Girardi et al. (2002) 
isochrone (solar metallicity) for a cluster age of $\log(t) = 8.8$. A
distance modulus of $V-M_V$ = 11.15 and reddening $E(B-V) = 0.10$~mag was
used.} 
\label{fig:cmd} 
\end{figure}

The physical parameters determined for the cluster NGC~6866 in the present study
are summarized in Table~3. We also compare our results with those given by
Kharchenko et al. (2005). In general, our values are in good agreement. 
However, we obtained a significantly smaller core radius and reddening than
Kharchenko et al. (2005).

\begin{table}
\caption{A comparison of the fundamental parameters of NGC~6866 from our
study and that of Kharchenko et al. (2005)}
\label{tab:funpar}
\begin{tabular} {lcc}
\hline
Cluster parameter          & Present study    & Kharchenko et al. (2005)  \\ \hline
$R_{core}$ (arcmin)& 2.0       & 4.2 \\
$R_{cluster}$ (arcmin)& 7.0    & 8.4 \\ 
Mean $E(B-V)$      & 0.10      & 0.17 \\
$V-M_{V}$          & 11.15     & 11.33\\
log(Age/yr)        & 8.8       & 8.68 \\ 
Distance (kpc)     & 1.47      & 1.45 \\  
\hline                                                                                
\end{tabular}                                                                       
\end{table}                                                                         

%
\section{Variable stars in the cluster NGC~6866}
\subsection{Selection criteria to identify variables}
After excluding observations taken under poor sky conditions, we considered only 718
photometric observations in $V$ and 47 in $I$ bands for the variability search. In order
to search variable stars in the target field, the time-series
$V$ band magnitude of all the stars are passed through the periodicity analysis
explained in the following section. A star was analysed for variability only
if it met the following criteria.
\begin{enumerate}
\item The star should not be within 10 pixels of the edge of any frame. 
\item The magnitude of the star should have a standard deviation less than three 
times the mean error in the corresponding magnitude bin.
\item  The stellar magnitude needs to be measured in at least 100 frames.
\end{enumerate} 
\subsection{Periodicity analysis}
We used the Lomb-Scargle periodogram (Lomb 1976, Scargle 1982) to estimate the period of a
variable star. This method computes the Fourier power spectrum by fitting sine and cosine
terms over a large number of frequencies in the given frequency range and is applicable to
unevenly sampled data. The period was derived using only the $V$-band data because of the
better sampling in this filter. In some cases, a few points were removed because either
they were deviated by more than 0.2~mag in two sequential observations or strongly deviated
from the mean magnitude. To search periodic variables, we have given a range of periods
between 0.01 days to 100 days as our total time length was about 100 nights in 2010. The
large range of period search was chosen because Lomb-Scargle method uses a period search in
the logarithmic time scale hence shallower at the longer periods. However, we have not
considered any period beyond 50 days, which is about half of the total observing span. 
We noticed that many spurious variables were found with periods which are harmonics or daily
aliases of each other. These stars were rejected. The only stars lie in the magnitude range
$V < 19.5$ were considered for variability search as photometric magnitudes have large errors
towards fainter end. Further, we have not considered 2008 observing data for the brighter stars
having $V<12.5$ mag which shows large uncertainty in the photometric measurements.
\begin{figure*}
\includegraphics[angle=0,width=1.0\textwidth, height=23cm]{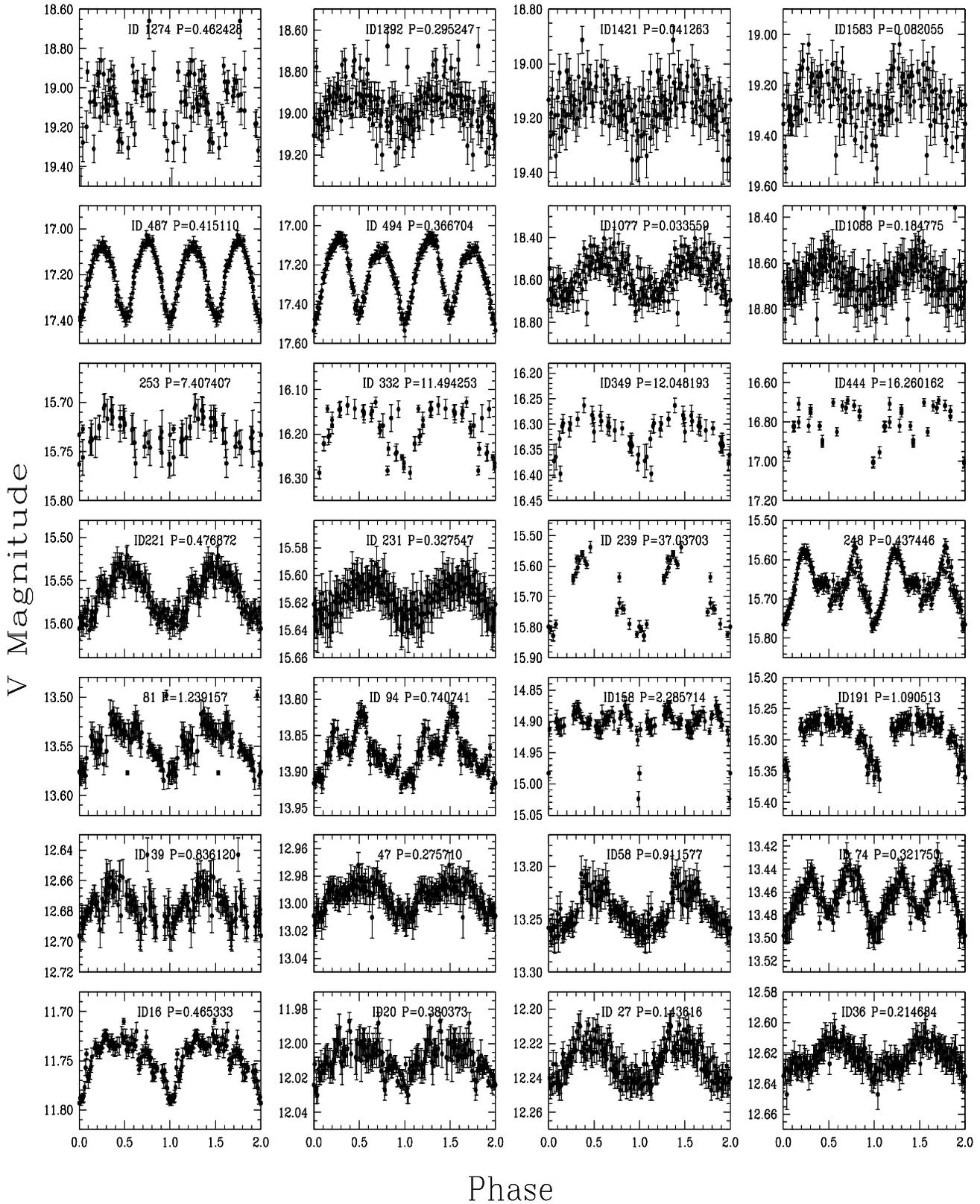}
\caption{$V$ band phase light curve for the 28 variable stars identified in the present study. Phase is
plotted twice and shifted in such a way that the light minimum falls near to zero phase. }
\label{fig:lc_mol}
\end{figure*} 

After these selection criteria, we visually inspected the light curves of
the remaining stars for periodic variation and found 28 variable stars.
A phased light curve was derived for each star using an estimated period. A 
phase-folded light curve, binned in intervals of 0.01 phase, was constructed 
by taking the average of the magnitudes in these phase bins. It was noticed that
some periodic variables, particularly in binary stars, the period-finding 
routine gives a `best-fit' period that actually corresponds to half the orbital 
period. To detect such variables, we inspected the light phased at twice the period
given by the period-finding algorithm.
For example, variable ID 248 gives a period of 0.217-day using the Lomb-Scargle 
algorithm, but by visual inspection we found that the true period is 0.434-d
because the light curve has two unequal minima. In this way, we found seven
stars which were better represented with twice the period given by the 
Lomb-Scargle periodogram.

\begin{table*}
\scriptsize
\caption{Details of the 28 periodic variables identified in the present study.
The variable stars identification are sorted in the order of increasing IDs. The columns
give RA, DEC, period, $V$ band mean magnitude, ($B-V)$ colour, amplitude of variation in
$V$ band, membership probabilities based on the spatial distribution, their position in
the CMD and proper motion, cluster membership, and their possible variability class. Last
column gives the characterization, identification number and period, if the star is already
identified in MOL09.}
\centering
\begin{tabular}{cccccccccclcc} \hline
 Star & RA         & DEC        &  Period   & $<V>$  & $B-V$&$\Delta$V& \multicolumn{4}{l}{Membership Probability} & \multicolumn{2}{l}{Classification based on} \\ 
 ID   & (J2000)    & (J2000)    &  (days)      & (mag)  & (mag)  & (mag)  & $P_{sp}$ & $P_{ph}$ & $P_{pm}$ &  Status& present study & MOL09 (ID, P) \\
\hline
0016 & 20:03:26.12 & 44:10:05.3 &  0.465333 & 11.747 & 0.34 & 0.064 & 0.28 & 1.00 & 0.66 & likely   & binary?	   & \\
0020 & 20:04:25.52 & 44:10:16.2 &  0.380373 & 12.003 & 0.32 & 0.032 & 0.19 & 1.00 & 0.16 & unlikely & binary?	   & \\
0027 & 20:03:47.13 & 44:09:25.7 &  0.143616 & 12.222 & 0.32 & 0.039 & 0.82 & 1.00 & 0.80 & member   & $\delta$ Sct & $\delta$ Sct (V3, 0.106$\pm$0.02)\\
0036 & 20:03:42.47 & 44:10:06.4 &  0.214684 & 12.623 & 0.32 & 0.035 & 0.70 & 0.47 & 0.84 & likely   & $\delta$ Sct & \\
0039 & 20:03:54.18 & 44:06:46.0 &  0.836120 & 12.677 & 0.25 & 0.035 & 0.60 & 1.00 & 0.81 & likely   & $\gamma$ Dor & $\gamma$ Dor (V12, 0.707)\\
0047 & 20:04:11.20 & 44:05:33.3 &  0.275710 & 12.995 & 0.31 & 0.041 & 0.27 & 1.00 & 0.55 & unlikely & ?            & $\delta$ Sct (V1, 0.066)\\
0058 & 20:03:31.61 & 44:07:59.8 &  0.911577 & 13.246 & 0.97 & 0.052 & 0.38 & 0.00 & 0.00 & field    & $\gamma$ Dor & \\
0074 & 20:03:34.93 & 44:14:50.1 &  0.321750 & 13.469 & 0.43 & 0.055 & 0.11 & 1.00 & 0.72 & likely   & Ell	   & W UMa      (V5, 0.321)\\
0081 & 20:03:27.93 & 44:09:19.1 &  1.239157 & 13.548 & 0.43 & 0.058 & 0.33 & 1.00 & 0.92 & likely   & $\gamma$ Dor & Irr      (V13)\\
0094 & 20:03:59.34 & 44:10:25.8 &  0.740741 & 13.878 & 0.43 & 0.087 & 0.82 & 1.00 & 0.77 & member   & $\gamma$ Dor & $\gamma$ Dor  (V11, 0.805)\\
0158 & 20:03:40.87 & 44:09:40.0 &  2.285714 & 14.902 & 0.68 & 0.052 & 0.66 & 1.00 & 0.75 & member   & EA	   & \\
0191 & 20:03:33.48 & 44:13:53.4 &  1.090513 & 15.285 & 0.67 & 0.085 & 0.19 & 1.00 & 0.67 & likely   & binary?	   & \\
0221 & 20:03:53.35 & 44:04:04.2 &  0.476872 & 15.564 & 0.83 & 0.079 & 0.21 & 1.00 & 0.86 & likely   & PV	   & \\
0231 & 20:03:48.28 & 44:10:56.6 &  0.327547 & 15.615 & 0.70 & 0.052 & 0.76 & 1.00 & 0.81 & member   & PV	   & \\
0239 & 20:04:19.00 & 44:07:05.8 & 37.037037 & 15.63  & 2.09 & 0.26 & 0.27 & 0.00 & 0.81 & unlikely  & semi-reg     & Irr	(V18)\\
0248 & 20:03:38.79 & 44:04:53.0 &  0.437446 & 15.66  & 0.82 & 0.21 & 0.22 & 1.00 & 0.74 & likely    & Ell	   & Ell	 (V9, 0.434)\\
0253 & 20:04:20.84 & 44:10:03.9 &  7.407407 & 15.73  & 1.18 & 0.06 & 0.31 & 0.00 & 0.32 & field     & Rot	   & Irr	 (V14)\\
0332 & 20:03:43.64 & 44:05:19.7 & 11.494253 & 16.19  & 1.44 & 0.14 & 0.33 & 0.00 & 0.18 & field     & semi-reg     & \\
0349 & 20:04:11.78 & 44:13:06.9 & 12.048193 & 16.31  & 1.41 & 0.10 & 0.32 & 0.00 & 0.97 & unlikely  & semi-reg     & \\
0444 & 20:04:21.25 & 44:15:40.6 & 16.260162 & 16.82  & 1.47 & 0.31 & 0.00 & 0.00 & 0.00 & field     & EB           & \\
0487 & 20:03:49.82 & 44:11:08.5 &  0.415110 & 17.21  & 0.93 & 0.34 & 0.76 & 1.00 & 0.85 & member    & W UMa	   & W UMa	(V7, 0.415)\\
0494 & 20:04:00.17 & 44:14:03.2 &  0.366704 & 17.26  & 1.03 & 0.44 & 0.34 & 1.00 & 0.00 & unlikely  & W UMa	   & W UMa	(V6, 0.366)\\
1077 & 20:04:13.87 & 44:03:45.8 &  0.033559 & 18.58  & 0.66 & 0.34 & 0.02 & 0.00 & 0.96 & unlikely  & HADS	   & \\
1088 & 20:03:56.20 & 44:12:49.9 &  0.184775 & 18.67  & 1.02 & 0.28 & 0.53 & 0.04 & 0.03 & field     & HADS	   & \\
1274 & 20:04:26.66 & 44:05:35.9 &  0.462428 & 19.05  & 1.42 & 0.46 & 0.00 & 1.00 &   -  & unlikely  & W UMa	   & W UMa	(V4, 0.262)\\
1292 & 20:03:41.23 & 44:12:17.9 &  0.295247 & 18.95  & 1.72 & 0.37 & 0.49 & 0.30 & 0.98 & unlikely  & HADS	   & \\
1421 & 20:03:41.08 & 44:08:47.4 &  0.041263 & 19.15  & 0.87 & 0.30 & 0.65 & 0.00 & 1.00 & unlikely  & HADS	   & \\
1583 & 20:03:58.70 & 44:11:33.5 &  0.082055 & 19.26  & 0.99 & 0.32 & 0.70 & 0.00 & 0.73 & likely    & HADS	   & \\
\hline
\end{tabular}
\end{table*}

Among the 28 periodic variables identified in the present study, 19 are newly detected.
Table~4 lists the identification number of variable stars, celestial coordinates, period,
and its membership probabilities. We also give the intensity-averaged mean magnitude
and the amplitude of brightness variation in $V$ band. The amplitude was estimated
as the difference between the median values of the upper and lower 15\%
magnitude values of the light curve (cf., Herbst et al. 2002; Messina et al. 2010).
In the last two columns, we give a possible class of variability from our analysis and
corresponding identification in MOL09 and period if the star has already been identified.
The $V$ band phased light curves of these variables are shown in Fig.~6.

Having the lesser number of data points, smaller amplitude of variation and larger
photometric error in $I$ band in comparison to $V$ band, we do not use $I$ band data
in our analysis. However, $I$ band phase light curves for all the 28 variables along
with their time-magnitude photometric data in $V$ and $I$ bands, are available in
electronic form.
In Fig.~7, we provide a finding  chart of a $\sim 13\times13$ arcmin field of the
cluster marking the variable stars.
\begin{figure*}
\includegraphics[width=18cm,height=18cm]{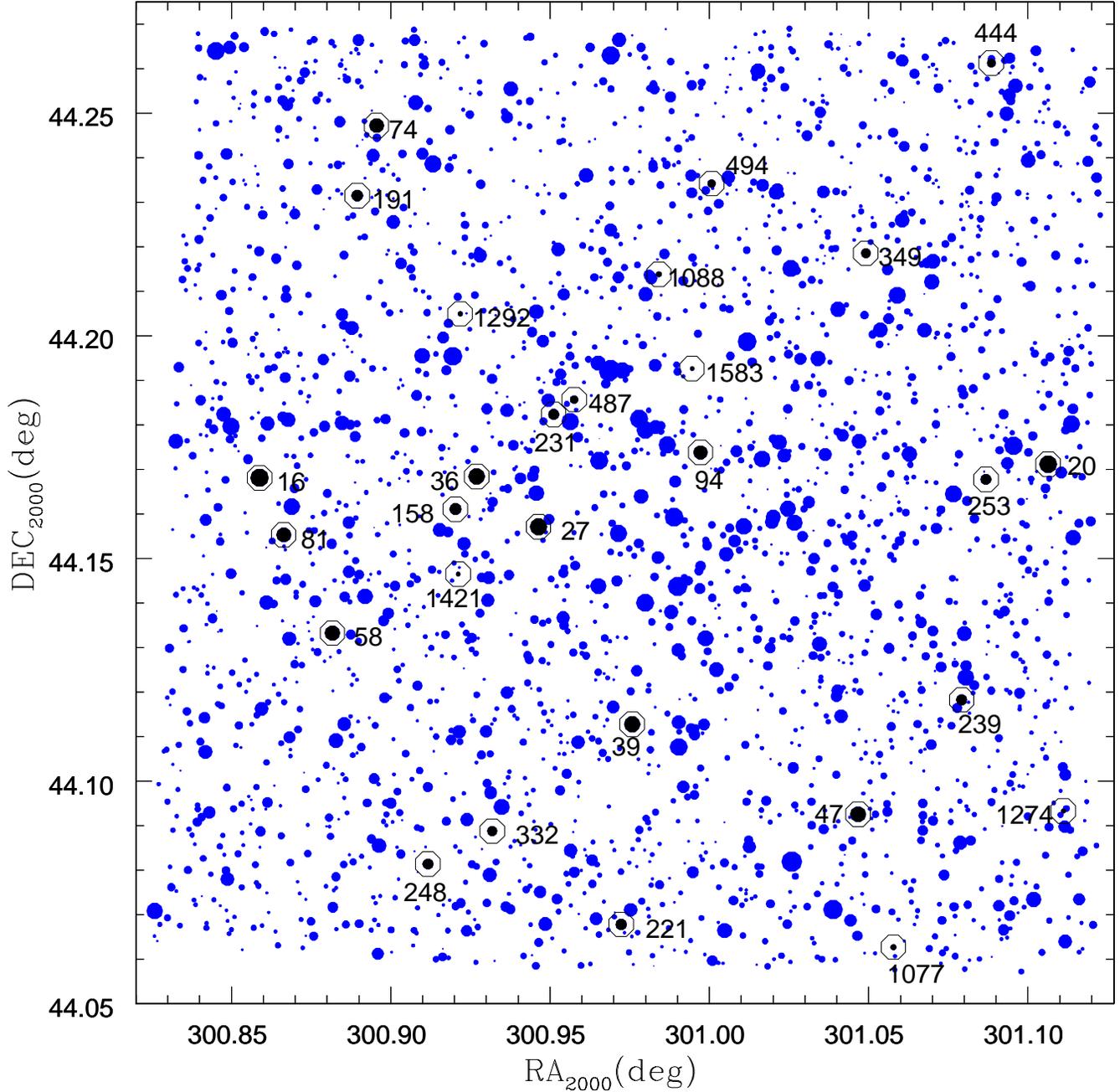} 
\caption{Finding chart for the variable stars in the field of NGC~6866. The size of
points signifies the magnitude of the stars. 28 periodic variables identified in the present study are additionally indicated with open circles. } 
\label{fig:fchart}
\end{figure*} 

%
\subsection{Comparison with other catalogues and previously known variables}
No time-series photometric observations of NGC~6866 have been published prior to MOL09. 
They detected 19 variables, 12 of which were periodic variables. Out of 19 variable stars, we are
able to find periodicity in 12 variables while 2 irregular variables V15 (ID 53) and V16 (ID 89)
in their list has also shown irregular variation in our data. Remaining five variables from
MOL09 are not included in our list of variable stars. Variable V2 (ID 30) is reported as
$\delta$~Scuti star with an amplitude variation of only 0.006 mag. We are unable to detect such
a low-amplitude variation in this bright ($V \sim 12.3$ mag) star owing to the uncertainties in
our data towards brighter end. Star V08 (ID 92) is heavily blended by two
very nearby stars brighter than this star. Though the light curve shows a significant flux variation
in our data but not an ellipsoidal in nature as reported by MOL09. Since this is a contaminated
star, its variability is highly susceptible to include it in our list of variable stars. The star V10
(ID 82) is listed as an eclipsing binary (EA) in their catalogue. For a period of 1.916~d, as reported
by MOL09 for this star, we should be able to confirm the variations because our observations extend
over a longer time span. However, we cannot confirm the eclipsing nature of this star from the present
data. Two stars in their catalogue, V17 and V19, which were classified as irregular variables lie close
to the edges of our frames. Since their reliable photometry could not be
determined in at least 100 frames, we exclude them from our list of variable stars.

It can be seen from the Table~4 that there is a significant difference between our periods and those of MOL09
for 5 variables. We therefore created phase-folded light curves using periods from  MOL09 for each
variable in common and compared them by eye with our phased light curves. We do not see any systematic 
periodic variation for these variables using their period apart from the $\delta$~Scuti star V3 which
could have either period. 
Variable V4 (ID 1274) is a faint star ($V \sim 19.1$ mag) which lies close to edge of our CCD chip in
most of the frames hence could be detected only in 111 frames. We therefore caution that the our
estimated period may not be very reliable due to its small number of data points.
\subsection{Cluster membership of the variable stars}
We assigned a probability of cluster membership to a star on the basis of its angular distance
from the cluster center, its location in the color-magnitude diagram and its proper motion with reference
to the mean proper motion of the cluster. The spatial probability ($P_{\rm sp}$), which gives the
position of the star in the field of the cluster, was  determined as $1-r_{i}/r_{c}$, where $r_{i}$ is
the angular distance of stars from the cluster center, and $r_{c}$ (7 arcmin) is the radius 
of the cluster.

The photometric probability ($P_{\rm ph}$) was computed with reference to the blue and
red limits in the $B-V$ vs $V$ diagram. The blue sequence was defined 
using empirical ZAMS colors (Schmidt-Kaler 1992) shifted in magnitude and color
by 11.15 and 0.10 mag respectively to visually match the cluster sequence. 
The red sequence was defined  by a shift of $-0.75$ mag in $V$ and a shift of 
0.042 in $B-V$ in order to account for unresolved MS binaries (Maeder 1974, 
Kharchenko 2004). Stars with $B-V$ colors lying within the binary sequence of the 
MS were assigned $P_{\rm ph}$ of 1 and are probable members of the cluster, while 
stars deviating along either direction were assigned a probability 
as $\exp [-0.5\times(\Delta(B-V)/\sigma_{(B-V)})^{2}]$, where $\Delta(B-V)$ is
the difference of color from blue or red color limits and $\sigma_{(B-V)}$
is the photometric error in color.

\begin{figure*}
\centering
\includegraphics[width=16cm]{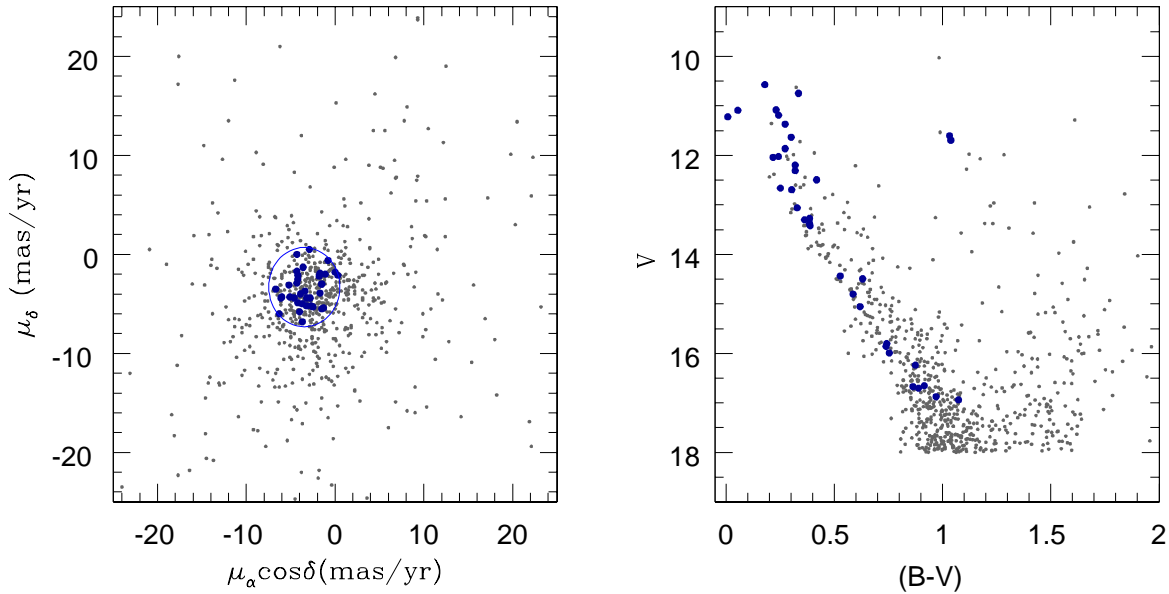}
\vspace{-7cm}
\caption{Proper motion distribution (left panel) and corresponding CMD (right panel)
in the field of NGC~6866. In both panels we show stars brighter than 18~mag in
$V$ by small dots, while stars used to estimate the mean proper motion are shown by 
filled circles. A circle of 4 mas radius is drawn on the left panel to mark stars 
having proper motion probability of 1. }
\label{fig:our_mol}
\end{figure*}

The proper motion probability ($P_{\rm pm}$) was calculated using the
catalogue of Roeser et al. (2010) which lists positions in ICRS system (accuracy 80 to 
300-mas) and absolute proper motion (accuracy 4 to 10-mas/yr) of about 900 million 
stars, derived from the optical USNOB1.0 and near-infrared 2MASS all sky catalogs. 
A cross-match of our photometry (2473 point sources) with that of Roeser et al. (2010) 
gave 1772 stars in common within a radius of 500 mas. For these stars, a scatter 
plot of the proper motion in right ascension ($\mu_{\alpha}$) versus that in 
declination ($\mu_{\delta}$) is plotted in Fig.~8. Unlike other nearby well studied
open clusters (see, e.g. Bellini et al. 2010), no clear trend separating
members and field stars is seen. We therefore computed the mean proper motion of the 
cluster with $V < 17$ mag, $Pph = 1$ and $Psp > 0.5$, which gives mean proper motions, 
$\bar{\mu}_x$ = $-3.47\pm0.41$ mas/yr and $\bar{\mu}_y$ = $-3.30\pm0.35$ mas/yr
\footnote{The mean proper motion of NGC~6866 is consistent with the value 
($-3.86\pm0.16,-4.63\pm0.17$) estimated by M09 who calculated an older
catalogue of Roeser et al (2008) which lists proper motion data of only brightest ($V < 14$ mag)
stars.}. The computations were done iteratively using a $\sigma$-clipping algorithm 
and the uncertainties denote $rms$ deviations. The $P_{ph}$ for each star was computed 
following Kharchenko et al (2004) as
$ \exp \left\{-0.25  \left[ (\mu_{\alpha} - \bar{\mu}_{\alpha})^{2}/ \sigma_{\alpha}^{2} +
(\mu_{\delta} - \bar{\mu}_{\delta})^{2}/ \sigma_{{\delta}}^{2} \right]  \right\} $,
where $\sigma_{\alpha}^{2} = \sigma_{\mu_{\alpha}}^{2}+\sigma_{\bar{\mu}_{\alpha}}^{2}$ and
$\sigma_{\delta}^{2} = \sigma_{\mu_{\delta}}^{2}+\sigma_{\bar{\mu}_{\delta}}^{2}$. We could
thus assign $P_{ph}$ for 1949 stars in common with the R10 catalog within a search radius of 
1 arcsec. The probabilities $P_{sp}$, $P_{ph}$ and $P_{pm}$ are listed in
Table~2. 

On the basis of spatial probability, stars found in the core of the cluster NGC~6866
($R_{core} = 2.0\pm0.5$ arcmin), for which $P_{\rm sp}$ is greater than 0.71$\pm$0.06, could be
the cluster member. The star having $P_{\rm ph}$=1.0 might belong to the cluster on the
basis of photometric probability and, stars which have $P_{\rm pm} > 60\%$ (1-$\sigma$)
are most probably belong to the cluster on the basis of proper motion criteria. Five
variable stars detected in the present study which satisfy all the three probability
criteria are considered as definite members of the cluster NGC~6866. 
Nine stars which satisfy two criteria are considered as likely members while other
9 stars those satisfy either only one criterion or have $P_{\rm pm} < 60\%$ are considered as
unlikely members. Five stars that do not follow any membership criteria
are considered as definite field stars. Membership probabilities and their status
as a cluster member are given in the Table~4.
\subsection{Characterization of the variable stars}
We assessed the classifications of variable stars by manually comparing
their phase-folded light curves to template light curves of different class of
variables. Our classification is primarily based on period, shape of the light curves and
location of the star in the CMD. In Fig.~9, we show the position of the variable stars in the
$(B-V)_0-M_V$ plane of the cluster. The intrinsic magnitude and color of the variable
stars are determined using the distance modulus $(m-M)_V$ = 11.15 mag and extinction
$E(B-V)$ = 0.10 mag as estimated in our study. The dashed line is the
ZAMS taken from Schmidt-Kaler (1982). In order to draw the classical instability
strip in the CMD, we transformed the theoretical instability strip boundaries of
Pamyatnysk et al. (2000) into our observational plane using  color-$T_{\rm
eff}$ relations from VandenBerg \& Clem (2003). The characterization of the variable 
stars detected in our study  is  summarized as follows:
\begin{figure*}
\centering
\includegraphics[angle=0, width=16.0cm]{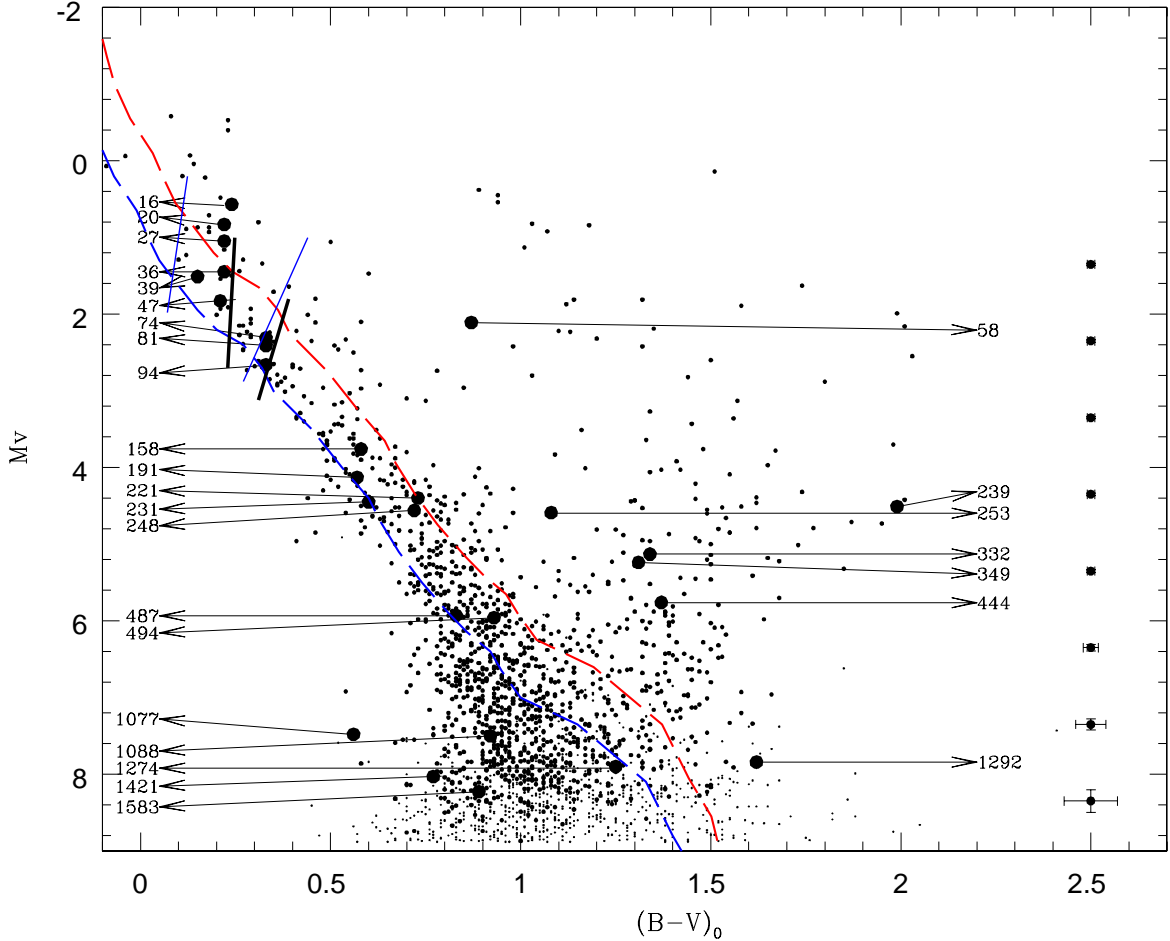}
\vspace{-3.2cm}
\caption{The $(B-V)_0$ vs $M_V$ CMD for NGC~6866. The thick
dashed line shown by blue line is the ZAMS corresponding to $V-M_V$ = 11.15 
and E(B-V) = 0.10 mag, while the red dashed line indicates a shift in magnitude 
and colour due to binarity. Positions of 28 periodic variables are also shown in
the CMD. The standard errors in colour and magnitude at various magnitude
ranges are given in the right corner. The thin solid lines represent the
instability strip for $\delta$-Sct stars while the thick solid
lines represent the same for $\gamma$-dor stars.}
\label{figure:cmd0}
\end{figure*}
%
\subsubsection{Previously identified variables}
ID 27 is classified as a $\delta$~Scuti by MOL09. In the CMD, this star is located within the
$\delta$~Scuti instability strip, justifying their classification. IDs 39 and 94 are reported
as $\gamma$-Dor variables. The location of ID 39 is towards the blue edge of the $\delta$~Scuti
instability strip, but its period suggests that it is a $\gamma$-Dor star. This makes it an
interesting target for the future observations. 
Star 81 is classified as irregular variable in MOL09. We found a periodicity of 1.24 day in the
present data, however, our observations do not give a uniform sampling at all the phases hence
period may be little uncertain. This star along with the star 94 fall in the $\gamma$-Dor
instability strip in the HR diagram which suggests that these stars belong to $\gamma$-Dor
class of variables. MOL09 classified stars 74, 487, 494 and 1274 as W Ursa
Majoris (W UMa) type variables. 
Our light curves support their classification for the stars 487, 494 and 1274, however, light
curve of the star 74 looks like an ellipsoidal variation.
Star 248 has an unusual light curve. This star is reported as an ellipsoidal variable but it
does not have the typical light curve of interacting binaries. It shows the light minima of equal
depth instead of the maxima. Such a strange light curve might arise if we view an interacting binary
in a highly-eccentric orbit close to pole-on. The variation in brightness is a result of the change
in sizes of the stars as they start approach periastron. Stars 239 and 253 are classified as
irregular variables by MOL09, but our phased light curves show clear periodic variability.
Star 239, which is an unlikely member of the cluster, is reddest ($B-V=2.09$, $V-I=3.44$) and
longest period ($\sim$ 37 day) variable identified in the present study. This star is most-likely
a background red giant and variability in these stars shows considerable periodicity in their light
variation, accompanied or sometimes interrupted by various irregularities as seen in its light curve.
We therefore classified this star as a semi-regular variable. Star 253 seems to be a rotating variable. 

Star 47 is reported as 0.066 day $\delta$-Sct variable (V1) by MOL09. Though this star falls
in the $\delta$-Sct instability strip in the CMD of our data, the shape of light curve is not a
typical $\delta$-Sct variation. Moreover, our study finds a period of 0.275 day and magnitude
variation of 0.041 mag for this star which is significantly higher in comparison of MOL09. While
proper motion probability for this star is reported as 72\% by MOL09, present study gives a
probability of only 55\%. Based on the spatial distribution, this star has only 27\% probability
of belonging to the cluster. Therefore, despite this star falling in the $\delta$-Sct instability
strip in the CMD of the cluster, present observations suggests that the star is unlikely a cluster
member and highly susceptible for being a $\delta$-Sct variable.
\subsubsection{New variables}
A total of 16 new variables were detected from the time-series data analysis of NGC~6866. 
We classify star 36, a likely member of the cluster, as $\delta$~Scuti variable as it is located in
the $\delta$~Scuti instability strip. Star 58, which is a field star, shows a light variation similar
to a $\gamma$-Dor variable.
Stars 16, 20 and 191 show similar light curves. They could be binary stars, though further data is needed
to ascertain their exact nature of variability.
Star 158 seems to be an eclipsing binary of the Algol (EA) type,  where the primary and secondary 
eclipses are clearly seen in the phased light curve.
Stars 221 and 231 show similar light curves. The periods and amplitudes are also similar. Star 221 is
a likely member, 231 is a confirmed member of the cluster based on their membership probabilities. Both
stars lie on or near the main sequence. These stars are placed in the group of pulsating variables (PV).
Stars 332 and 349 look like a semi-regular or long period variable (LPV). 
Star 444 has a periodicity of $\sim$ 16.3 days and display signature of primary and secondary eclipses.
More observations will be needed to see light variations at all the phases for this long period
variable, nevertheless, we classify this star as $\beta$~Lyr (EB) type binary based on present observations. 
Stars 1077, 1088, 1292, 1421 and 1583 have periods similar to $\delta$~Scuti stars, but have significantly
large amplitudes, suggesting that these stars could be high-amplitude $\delta$-Sct stars (HADS). 

Present study along with MOL09 classified stars 487, 494 and 1274 as W~UMa type variable which are
known to obey a PLC relationship between the period, color and absolute brightness (Rucinski 2004) as 
follows: 
$$M_V = -4.44 log(P) + 3.02(B-V)_0 + 0.12 ~~~~~~~~~~~ \sigma = 0.25~mag$$
\noindent 
where $(B-V)_0$ is the intrinsic color index and $P$ is the orbital period in days. We derived the apparent
distance modulus $(m-M)_V$ of each W~UMa star from the absolute magnitude $M_V$ determined using the above
relation and their mean $V$ magnitudes. A reddening of $E(B-V)=0.10$ was assumed.
Of the three W~UMa variables, star 487 is the most interesting. It has a very high proper motion
probability of 0.85, spatial probability of 0.76 and photometric probability of 1.0 and hence 
considered as a confirmed member of the cluster. However, $(m-M)_V$ of this star 
is estimated as 12.53 mag from the W~UMa PLC relationship. This is 1.38 mag fainter than the apparent
distance modulus of the cluster. Since the period and colour of this star is quite robust, a large
value of $(m-M)_V$ suggests that either this star is not a cluster member or it does not belong to
class of W~UMa variables hence not following the above relationship. This star needs further
attention to ascertain its true membership and nature of variability.
Other two stars 494 and 1274 has zero proper motion and no proper motion information available 
respectively. Though both of them fall in the CMD, their spatial position in the target field is
quite far from the cluster center. Furthermore, stars 494 and 1274 are fainter by 0.94 and 2.03 mag
respectively than the estimated distance modulus of the cluster. This 
makes these two W~UMa stars unlikely members of the cluster. 
\section{Summary}
The work presented here is the first paper of our series of papers on detection
and characterization of variable stars in young and intermediate age open clusters. 
In this work we have presented a search for variable stars in the cluster NGC~6866. 
We found 28 variables in the period range 0.03-day ($\sim$ 48-mins) to
37-day and confirm the irregularity in the light curves of other two irregular variables
reported by MOL09. Among them, 16 are newly-discovered periodic variables. We also determined
period for 3 other variables which were reported as irregulars in MOL09. Prior to
this study, the shortest period variable detected in NGC~6866 was a 13~mag
$\delta$-Sct star with a period of 1.6-hr. Since we have carried out continuous
observations in the $V$ band for more than 4 hours on 3 nights, this has enabled
us to find variable star with period as short as 48 min. In the field of NGC~6866, we
did not find any variable star which 
shows variations in excess of 0.5 mag. In comparison with previous studies,
we report more than twice the number of periodic variables in the field of NGC~6866.

In the present study, we have analysed membership of the stars on the basis of their 
distance from the cluster center, positions on the CMD, and their proper 
motions, wherever available. The membership probabilities of the stars on the basis 
of all the three criteria were estimated. Our analysis of 28 periodic variables
suggests that 14 variables are either confirmed members or 
likely members of the cluster, while remaining 14 variables are either field stars or
unlikely members of the cluster. Based on the shape and period of the light variations,
together with the colour and amplitude, we found several $\delta$~Scuti, $\gamma$-Dor,
rotational variable and eclipsing binaries. Few variables could not be classified with
precision and more photometric observations are needed to ascertain the exact nature
of these stars. Further multi-colour photometry and spectroscopic observations of these
stars will help to determine their parameters such as masses and radii. 

We also provide calibrated $UBVRI$ photometry of 2473 stars down to
$V \sim 21.5$ mag. These data were used to determine the physical parameters of the 
cluster and derive the cluster radius from the stellar density profile and the 
extinction, distance and age from the  colour-colour and colour-magnitude diagrams. 
The basic parameters of the cluster NGC~6866 is obtained through isochrone
fitting, giving $\log(t) = 8.8$~yr, a distance modulus of $(m-M)_0$ = 10.84 mag, 
and extinction $E(B-V)$ = 0.10 mag. The radial distribution of the stellar surface 
density indicates that the core and cluster radius is extended up to about 2 and 
7 arcmin respectively with a peak density of $5.7\pm0.7$ star/arcmin$^2$ at 
the cluster center. 

\section*{Acknowledgments}
We are thankful to J. Molenda-\.{Z}akowicz for providing photometric data of their published
variables. Thanks to Rama Kant, N. K. Chakradhari, Manoj Patel and Vindor Kumar for their
observational support. SJ acknowledge the grant received 
under the Indo-South Africa Science and Technology Cooperation INT/SAFR/P-3(3)2009) funded
by Departments of Science and Technology of the Indian and South African Governments.

\label{lastpage}

\end{document}